\documentclass[letter]{aa} 

\usepackage{txfonts}

\begin{document}

\titlerunning{Novel distance indicator for GRBs associated with SNe}

\title{Novel distance indicator for Gamma-Ray Bursts associated with Supernovae}

\authorrunning{G. B. Pisani et al.}

\author{G. B. Pisani\inst{1,3},  L. Izzo\inst{1,2}, R. Ruffini\inst{1,2}, C. L. Bianco\inst{1,2}, M. Muccino\inst{1}, A. V. Penacchioni\inst{1,3}, J. A. Rueda\inst{1,2}, Y. Wang\inst{1}}

\institute{\inst{1}Dipartimento di Fisica, Sapienza Universit\'a di Roma and ICRA, Piazzale Aldo Moro 5, I-00185 Roma, Italy,\\
\inst{2}ICRANet, Piazza della Repubblica 10, I-65122 Pescara, Italy,\\
\inst{3}Universit\'e de Nice Sophia Antipolis, Nice, CEDEX 2, Grand Chateau Parc Valrose.}

\offprints{\email{gb.pisani@icranet.org}}

\abstract
{
It has been proposed that the temporal coincidence of a gamma-ray burst (GRB) and a type Ib/c supernova (SN) can be explained with the concept of induced gravitational collapse (IGC), induced by the matter ejected from an SN Ib/c accreting onto a neutron star (NS). The NS is expected to reach the critical mass necessary for it to collapse to a black hole (BH) and emit a GRB. We found a standard luminosity light curve behavior in the late-time X-ray emission of this subclass of GRBs.
}
{
We test if this standard behavior in the luminosity found in this subclass of GRBs can become a redshift estimator of these sources.
}
{
We selected a sample of GRBs that belong to this subclass of IGC GRBs associated to an SN (IGC GRB-SN sources). These sources have an isotropic energy $E_{iso} > 10^{52}$ erg and their cosmological redshifts are in the range of $z = 0.49\,$--$\,1.261$. We focused on the corresponding X-ray luminosity light curves. 
}
{
We find that all GRBs of the sample with measured redshift present a standard luminosity late-time light curve in the $0.3\,$--$\,10$ keV rest-frame energy range. We used these results to estimate the GRB redshift of the sample without a measured redshift, and found results consistent with other possible redshift indicators.
}
{
The standard late-time X-ray luminosity light curve of all GRBs of the sample shows a common physical mechanism in this particular phase of the X-ray emission, possibly related to the creation of the NS from the SN process. This scaling law moreover represents strong evidence of very low or even absent beaming in this late phase of the X-ray afterglow emission process. This could be a fundamental tool for estimating the redshift of GRBs that belong to this subclass of events. We are currently expanding this subclass of GRBs to further verify the universal validity of this new redshift estimation method.
}


\date{}

\maketitle

Recently, \citet{Ruffini2001c,Ruffini2007b} proposed that the temporal coincidence of some gamma-ray bursts (GRBs) and a type Ib/c supernovae (SNe) can be explained with the concept of induced gravitational collapse (IGC) of a neutron star (NS) to a black hole (BH) induced by accretion of matter ejected by the SN Ib/c. More recently, this concept has been extended, including a precise description of the progenitor system of such GRB-SN systems \citep{Rueda2012}.

The main new result presented here is that the IGC GRB-SN class shows a standard late X-ray luminosity light curve in the common energy range $0.3\,$--$\,10$ keV \citep{RuffiniMG13}.

The prototype is GRB 090618 \citep{TEXAS,Izzo2012,Izzo2012b} at redshift $z=0.54$, where four different emission episodes have been identified. 

Episode 1, corresponding to the SN onset, has been observed to have thermal as well as power-law emission. The thermal emission changes in time following a precise power-law behavior \citep{Izzo2012,Penacchioni2012,Ana2013}.

Episode 2 follows and in the IGC model corresponds to the GRB emission coincident with the BH formation. The characteristic parameters of the GRB, including baryon load, the Lorentz factor, and the nature of the circumburst medium (CBM), have been computed \citep{Izzo2012,Penacchioni2012,Ana2013}. 

Episode 3 is characterized in the X-ray light curve by a shallow phase (a plateau) followed by a final steeper decay. Typically, it is observed in the range $10^2\,$--$\,10^6$ s after the GRB trigger. 

Episode 4 occurs after a time of about ten days in the cosmological rest-frame, corresponding to the SN emission due to the Ni decay \citep[see][for a complete review]{Arnett}. This emission is clearly observed in GRB 090618 during the late optical GRB afterglow emission.

Here we analyze the X-ray emission of a sample of eight GRBs with $E_{iso} \geq 10^{52}$ erg that satisfy at least one of the following three requirements:
\begin{itemize}
\item there is a double emission episode in the prompt emission: Episode 1, with a decaying thermal feature, and Episode 2, a canonical GRB, as in GRB 090618 \citep{Izzo2012}, GRB 101023 \citep{Penacchioni2012}, and in GRB 110709B \citep{Ana2013};
\item there is a shallow phase followed by a final steeper decay in the X-ray light curve: Episode 3;
\item an SN is detected after about ten days from the GRB trigger in the rest-frame: Episode 4.
\end{itemize}

\begin{table}
\centering
\begin{tabular}{l c c}
\hline\hline
GRB  & $z$ & $E_{iso} (erg)$ \\
\hline 
GRB 060729 & $0.54$ & $1.6 \times 10^{52}$ \\
GRB 061007 & $1.261$ & $1.0 \times 10^{54}$ \\
GRB 080319B & $0.937$ & $1.3 \times 10^{54}$ \\
GRB 090618 & $0.54$ & $2.9 \times 10^{53}$ \\
GRB 091127 & $0.49$ & $1.1 \times 10^{52}$ \\
GRB 111228 & $0.713$ & $2.4 \times 10^{52}$ \\
\hline
GRB 101023 & $0.9^*$ & $1.8 \times 10^{53}$ \\
GRB 110709B & $0.75^*$ & $1.7 \times 10^{53}$ \\
\hline\\
\end{tabular}
\caption{GRB sample considered in this work. The redshifts of GRB 101023 and GRB 110709B, which are marked with an asterisk, were deduced theoretically by using the method outlined here \citep{Penacchioni2012} and the corresponding isotropic energy computed by assuming these redshifts.}
\label{table1}
\end{table}

\begin{figure*}
\centering
\includegraphics[width=0.78\hsize,clip]{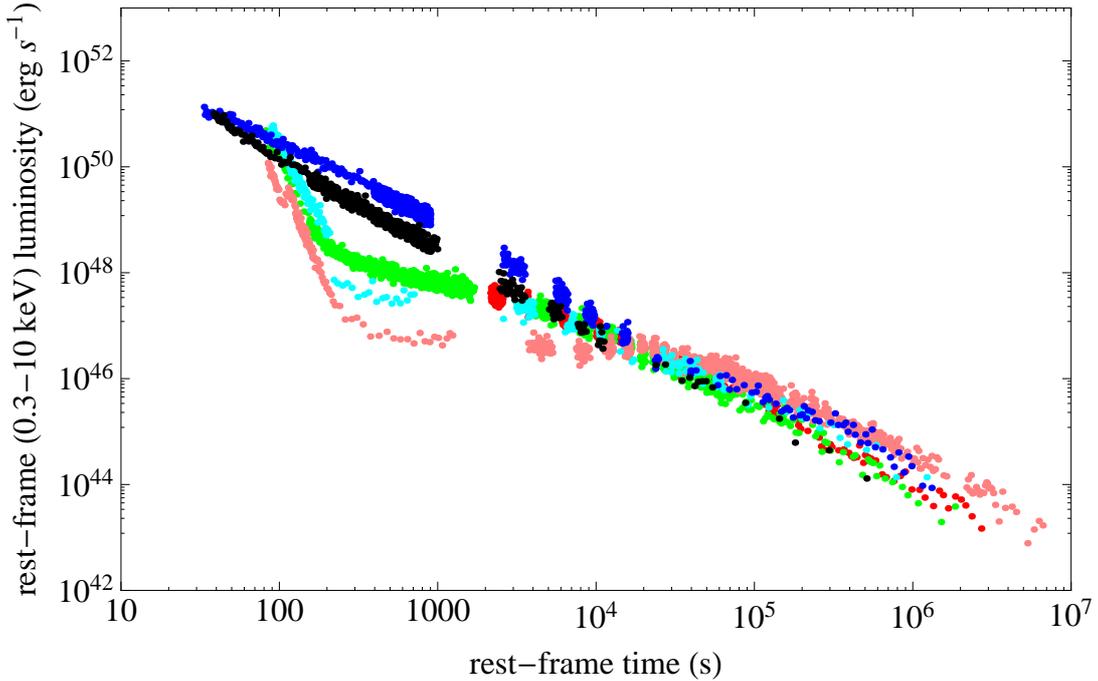}
\caption{X-ray luminosity light curves of the six GRBs with measured redshift in the $0.3\,$--$\,10$ keV rest-frame energy range: in pink GRB 060729, $z=0.54$; black GRB 061007, $z=1.261$; blue GRB 080319B, $z=0.937$; green GRB 090618, $z=0.54$, red GRB 091127, $z=0.49$, and in cyan GRB 111228, $z=0.713$.}\label{fig:sample}
\end{figure*}

We found eight GRBs that satisfy our requirements (see Table \ref{table1}).

\textit{GRB 060729}. In this source an SN bump was observed in the optical GRB afterglow \citep{Cano2011}. It is at the same redshift $z=0.54$ as GRB 090618 and shows a small precursor plus a main event in the prompt light curve and a peculiar prolonged duration for the X-ray afterglow. The isotropic energy emitted in this burst is $E_{iso} = 1.6 \times 10^{52}$ erg \citep{Grupe2007b}.

\textit{GRB 061007}. This GRB has no associated SN but is characterized by a precursor with a clear evolving thermal emission \citep{Larsson2011}. With an energetic $E_{iso}$ of $1.0 \times 10^{54}$ erg \citep{GCN5722} at $z=1.261$, it is the farthest GRB in our sample. The large distance makes the detection of an SN from this GRB difficult.

\textit{GRB 080319B}. A tentative SN was reported also for GRB 080319B, well-known as the naked-eye GRB, whose prompt emission also shows a possible double-emission episode \citep{Kann2008}. Its measured redshift is $z=0.937$. This is one of the most energetic GRB, with $E_{iso} = 1.3 \times 10^{54}$ erg \citep{GCN7482}, and its X-ray light curve is well fit by a simple decaying power-law.

\textit{GRB 090618}. This GRB is the prototype of the IGC GRB-SN subclass. Its prompt emission shows a clear Episode 1 plus Episode 2 structure in light curve and spectrum. The measured redshift is $z=0.54$ and the $E_{iso} = 2.9 \times 10^{53}$ erg \citep{Izzo2012}. There is a clear optical bump, about ten days of rest-frame time after the GRB trigger, in the afterglow light curve of GRB 090618 that is associated with the SN emission \citep{Cano2011}. The characteristic parameters of this GRB, including the baryon load ($B=1.98 \times 10^{-3}$), the Lorentz factor at the trasparency ($\Gamma_{tr}=495$), and the nature of the CBM ($\langle n_{CBM} \rangle=0.6 \, part/cm^3$), have been estimated previously \citep{Izzo2012}.

\textit{GRB 091127}. GRB 091127 is associated with SN 2009nz at a redshift of $z=0.49$ \citep{Cobb2010}. The $E_{iso}$ for this burst is $1.1 \times 10^{52}$ erg \citep{GCN10204}.

\textit{GRB 111228}. An SN feature was also reported in the literature for GRB 111228 \citep{DAvanzo2012}, which shows a multiply peaked prompt light curve in the Fermi-GBM data. The measured redshift of this GRB is $z=0.713$, its $E_{iso} = 2.4 \times 10^{52}$ erg \citep{GCN12744}, and a dedicated analysis of this GRB will be presented elsewhere. The detection of an SN in GRB 111228 is debated, since the subsequent optical bump has the same flux as the host galaxy of the source, but SN features were observed in the differential photometry between the last epochs of observations, where a transient component was detected that was unrelated to the afterglow, and was consequently attributed to the SN. 

\textit{GRB 101023}. This GRB shows clear Episode 1 and Episode 2 emissions in the prompt light curve and spectrum, but there is no detection of an SN and no measured redshift because of the lack of optical observations at late times. We estimated the redshift of this source at $z=0.9$ by analogy to the late X-ray afterglow decay observed in the six GRBs with a measured redshift. This leads to an estimate of $E_{iso} = 1.8 \times 10^{53}$ erg, a baryon load of $B=3.8 \times 10^{-3}$, a Lorentz factor at transparency of $\Gamma_{tr}=260$, and an average density for the CBM of ($\langle n_{CBM} \rangle \approx 16 \, part/cm^3$ \citep{Penacchioni2012}.

\textit{GRB 110709B}. Like GRB 101023, this GRB shows clear Episode 1 plus Episode 2 emission in the prompt light curve and spectrum, but there is no detection of an SN. This can be explained by the fact that it is a dark GRB, whose emission is strongly influenced by absorption. Particularly interesting is the detection of clear radio emission \citep{Zauderer2012}. There is no measurement for the redshift but, as for GRB 101023, we estimated it to be $z=0.75$ by analogy to the late X-ray afterglow decay observed in the six GRBs with measured redshifts. This leads to an estimate of an isotropic energy of $E_{iso} = 1.7 \times 10^{53}$ erg, a baryon load of $B=5.7 \times 10^{-3}$, a Lorentz factor at the trasparency of $\Gamma_{tr}=174$, and an average density of the CBM of $\langle n_{CBM} \rangle \approx 76 part/cm^3$ \citep{Ana2013}.

We focused the analysis of all available XRT data of these sources. Characteristically, XRT follow-up starts only about 100 seconds after the BAT trigger (typical repointing time of Swift after the BAT trigger). Because the behavior was similar in all sources, we compared the analyzed XRT luminosity light curve $L_{rf}$ for the six GRBs with measured redshifts in the common rest-frame energy range $0.3\,$--$\,10$ keV. As a first step we converted the observed XRT flux $f_{obs}$ to one in the $0.3\,$--$\,10$ keV rest-frame energy range. In the detector frame, the $0.3\,$--$\,10$ keV rest-frame energy range becomes $[0.3/(1+z)]\,$--$\,[10/(1+z)]$ keV, where $z$ is the redshift of the GRB. We assumed a simple power-law function as the best fit for the spectral energy distribution of the XRT data\footnote{http://www.swift.ac.uk/}:
\begin{equation}
\frac{dN}{dA\,dt\,dE} \propto E^{-\gamma}\,.
\label{spettro_pl}
\end{equation}
We can then write the flux light curve, $f_{rf}$, in the $0.3\,$--$\,10$ keV rest-frame energy range as
\begin{equation}
f_{rf} = f_{obs} \frac{\int_{\frac{0.3\,keV}{1+z}}^{\frac{10\,keV}{1+z}}E^{-\gamma}dE}{\int_{0.3\,keV}^{10\,keV}E^{-\gamma}dE} = f_{obs} (1+z)^{\gamma-1}\,.
\label{flusso_1}
\end{equation}
Then, we have to multiply $f_{rf}$ by the luminosity distance to derive $L_{rf}$:
\begin{equation}
L_{rf} = 4 \, \pi \, d_l^2(z) f_{rf}\,,
\label{luminosity}
\end{equation}
where we assume a standard cosmological $\Lambda$CDM model with $\Omega_m = 0.27$ and $\Omega_{\Lambda}=0.73$. Clearly, this luminosity must be plotted as a function of the rest-frame time $t_{rf}$, namely
\begin{equation}
\label{time_correction}
t_{rf} = \frac{t_{obs}}{1+z}\,.
\end{equation}

\begin{figure*}
\centering
\includegraphics[width=0.49\hsize,clip]{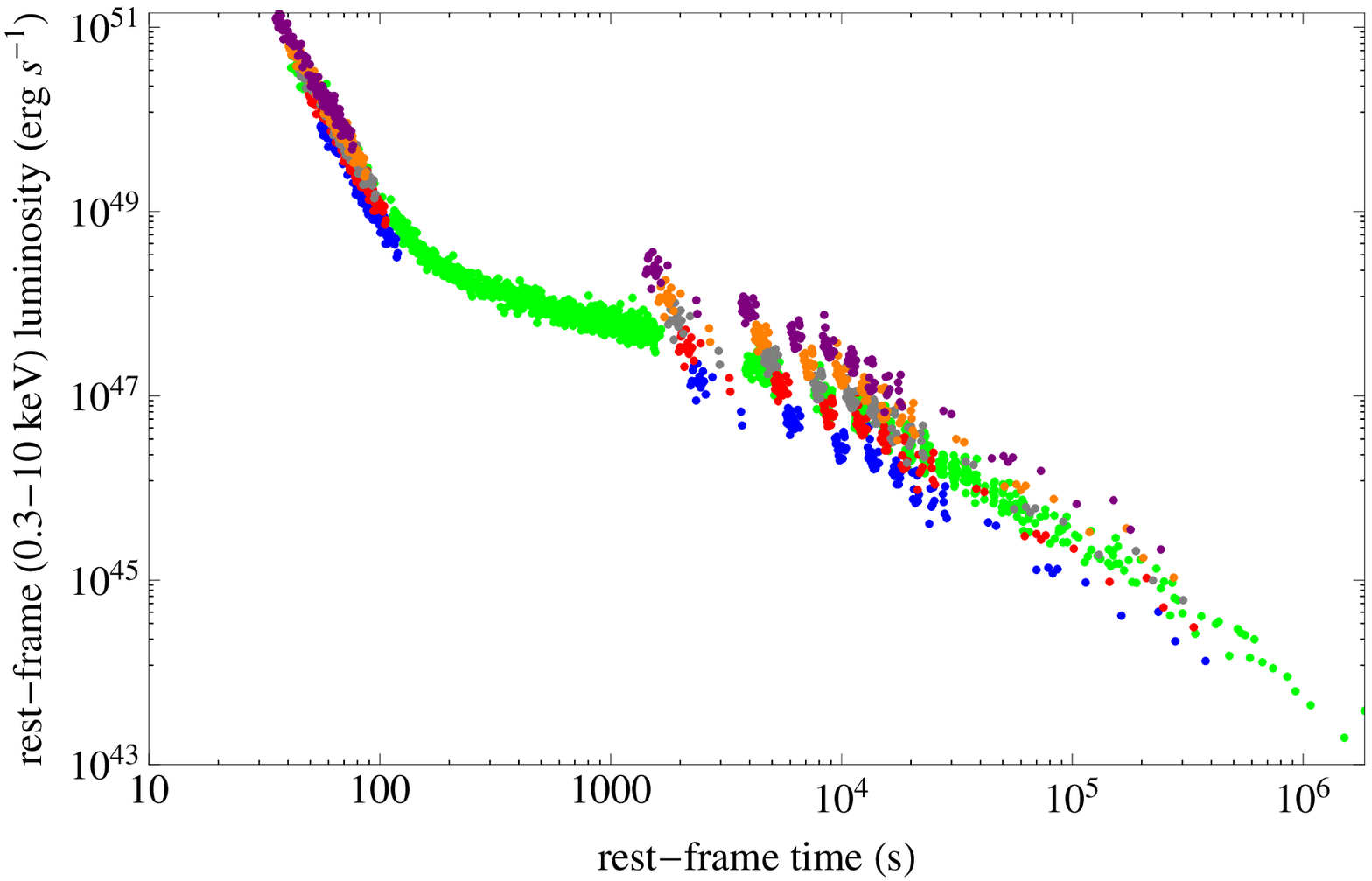}
\includegraphics[width=0.49\hsize,clip]{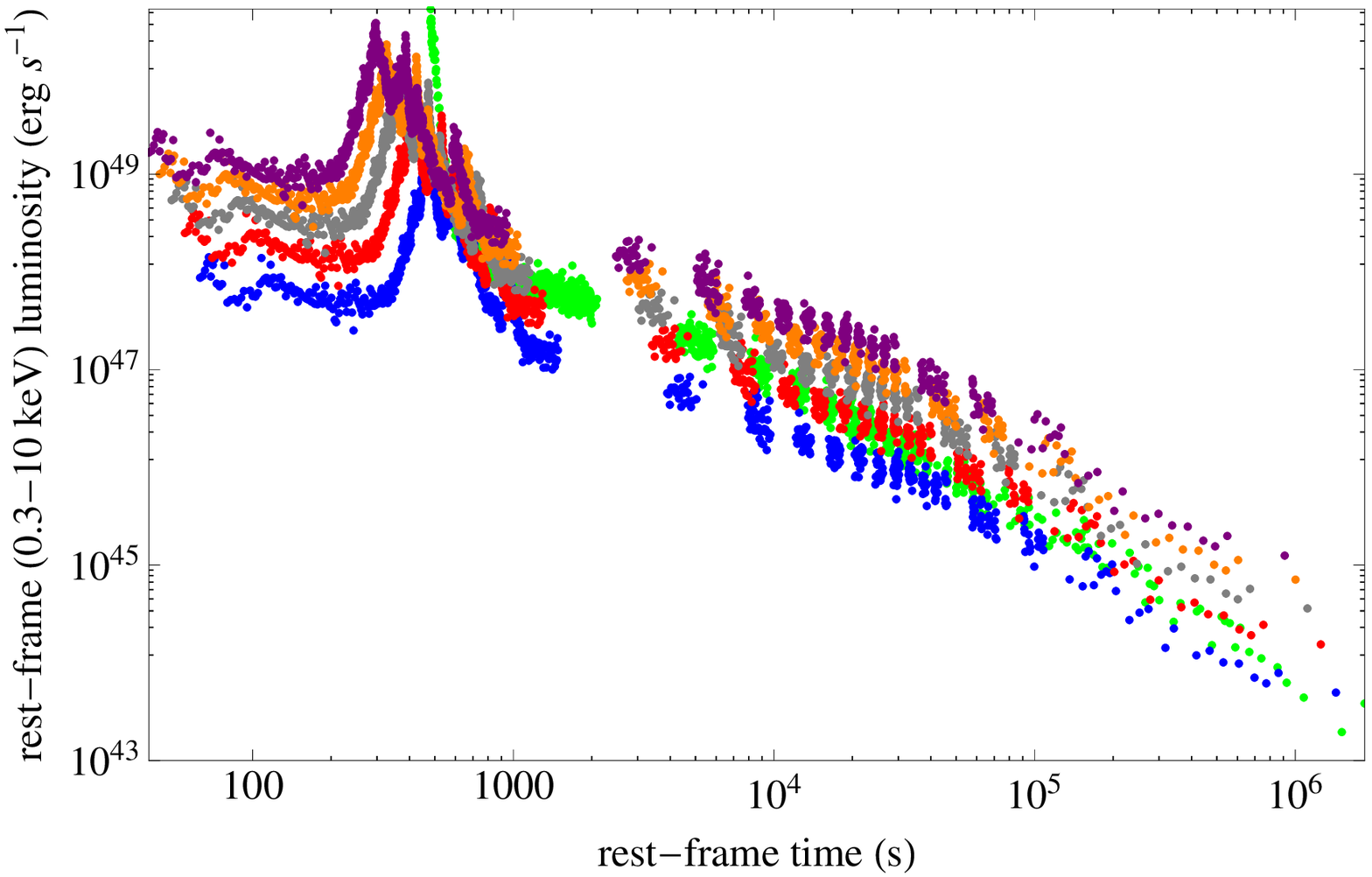}
\caption{In green we show the rest-frame X-ray luminosity light curve of GRB 090618 in the $0.3\,$--$\,10$ keV energy range in comparison with the one of GRB 101023 (left) and GRB 110709B (right), computed for different hypothetical redshifts: respectively, from blue to purple: $z=0.6, 0.8, 1.0, 1.2, 1.5$ (left) and $z=0.4, 0.6, 0.8, 1.0, 1.2$ (right). The overlapping at late time of the two X-ray luminosity light curves is obtained for a redshift of $z=0.9$ (left) and $z=0.75$ (right). For further details see \citet{Penacchioni2012,Ana2013}.}
\label{fig:101023}
\end{figure*}

The X-ray luminosity light curves of the six GRBs with measured redshifts in the $0.3$--$10$ keV rest-frame energy band are plotted in Fig. \ref{fig:sample}. What is most striking is that these six GRBs, with redshifts in the range $0.49\,$--$\,1.261$, show a remarkably common behavior of the late X-ray afterglow luminosity light curves (Episode 3), despite their very different prompt emissions (Episode 1 and 2) and energetics spanning more than two orders of magnitude. The common behavior starts between $10^4\,$--$\,10^5$ s after the trigger and continues until the emission falls below the XRT threshold. This standard behavior of Episode 3 represents a strong evidence of very low or even absent beaming in this particular phase of the X-ray afterglow emission process. We have proposed that this late-time X-ray emission in Episode 3 is related to the process of the SN explosion within the IGC scenario, possibly emitted by the newly born NS, and not by the GRB itself \citep{Negreiros2012}. This scaling law, when confirmed in sources with Episode 1 plus Episode 2 emissions, offers a powerful tool for estimating the redshift of GRBs that belong to this subclass of events.
As an example, Fig.~\ref{fig:101023} plots the rest-frame X-ray luminosity (0.3 - 10 keV) light curve of GRB 090618 (considered the prototype of the common behavior shown in Fig. \ref{fig:sample}) with the rest-frame X-ray luminosity light curves of GRB 110709B estimated for selected values of its redshifts $z=0.4, 0.6, 0.8, 1.0, and 1.2$, and similarly the corresponding analysis for GRB 101023 for redshifts $z=0.6, 0.8, 1.0, 1.2, and 1.5$. This shows that GRB 101023 should have been located at $z \sim 0.9$ and GRB 110709B at $z \sim 0.75$. These redshift estimates are within the range expected using the Amati relation as shown in \citet{Penacchioni2012,Ana2013}. This is an important independent validity confirmation for this new redshift estimator we are proposing for the family of IGC GRB-SN systems. We stress, however, that the redshift was determined assuming the validity of the standard $\Lambda$CDM cosmological model for sources with redshift in the range $z=0.49\,$--$\,1.216$. We are currently testing the validity of this assumption for sources at higher cosmological redshifts.

\begin{figure}
\centering
\includegraphics[width=\hsize,clip]{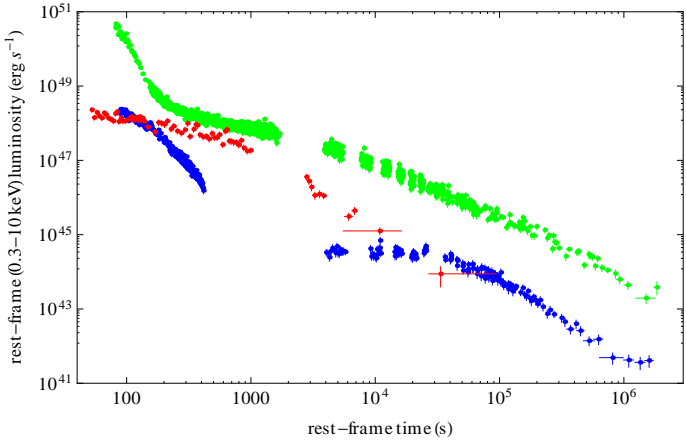}
\caption{X-ray luminosity light curves of GRB 090618 (green), GRB 060614 (blue), and GRB 090510 (red) in the $0.3\,$--$\,10$ keV rest-frame energy range.}\label{fig:3families}
\end{figure}

Before concluding, it is appropriate to recall once again that we assumed that these binary systems give rise to the IGC GRB-SN sources, which are a subclass of all GRBs. Their special binary nature is very different from that corresponding to the genuine short GRBs, for instance.
For these the progenitors are thought by many to be binary neutron stars and there is no expected observable afterglow emission \citep[see e.g. the case of GRB 090227B presented in][]{Marco2013}. They are also different from disguised short GRBs, which again may originate from binary systems drifting to the galactic halo \citep{Bernardini2007,Caito2009,Caito2010,deBarros2011}. In particular, they may differ from GRB 060614, where there is strong evidence that it has no associated SN \citep{DellaValle2006b,Gal-Yam2006,Gehrels2006}.
We were also able to show explicitly that the X-ray luminosity light curve of the IGC GRB-SN prototype, GRB 090618, is drastically different both from that of GRB 060614 and that of GRB 090510, which may be an example of a disguised short GRB that may have instead exploded in a very high density region \citep{Marco2013}, see Fig. \ref{fig:3families}. 
In all the above examples we have considered very energetic sources ($E_{iso} \geq 10^{52}$ erg). Less energetic GRB-SN sources, e.g. GRB 980425 \citep{2000ApJ...536..778P}, also show a late X-ray emission different from the typical emission of the IGC GRB-SN sources, and we will discuss this matter elsewhere. 

We presented a sample of IGC GRB-SN systems with a standard late-time ($10^4\,$--$\,10^5$ s after the trigger) X-ray luminosity light curve in the $0.3\,$--$\,10$ keV rest-frame energy band. This standard behavior points to a common physical origin of this emission, possibly related to a newly born NS out of the SN event \citep{Negreiros2012}. This scaling law can provide a new distance indicator for this subclass of GRBs, allowing one to predict the redshift of the source as well as the presence of an associated SN.

We are currently testing the predictive power of our results on three different observational scenarios for sources of the IGC GRB-SN subclass:
\begin{itemize}
\item GRBs at high redshift. We are able to predict the existence of an SN in these systems, which is expected to emerge after $t \sim 10 \, (1+z)$ days, the canonical time sequence of an SN explosion. This offers a new challenge to detect SNe at high redshift, e.g., by observing radio emission \citep{Ana2013};
\item for GRBs with $z\leq1$  we can indicate in advance from the X-ray luminosity light curve observed by XRT the expected time for the observations of an SN and alert direct observations from ground- and space-based telescopes; 
\item as we showed here, we can infer the redshift of GRBs in the same way we did for GRB 110709B and GRB 101023A.
\end{itemize}

We are currently expanding the sample to increase the statistical validity of our approach and its cosmological implications.

\begin{acknowledgements}
We are grateful to the anonymous referee for important remarks that have improved the presentation of our paper. We also thank the \textit{Swift} team for the support. This work made use of data supplied by the UK \textit{Swift} Data Center at the University of Leicester. G.B. Pisani and A.V. Penacchioni are supported by the Erasmus Mundus Joint Doctorate Program by Grant Numbers 2011-1640 and 2010-1816, respectively, from the EACEA of the European Commission.
\end{acknowledgements}

\end{document}